\documentclass[%
 reprint, 
 prl,
 twocolumn,
superscriptaddress,
groupedaddress,
 amsmath,amssymb,
 aps, 
]{revtex4-2}

\usepackage{graphicx}
\usepackage{dcolumn}
\usepackage{bm}

\begin{document}


\title{\textbf{Shedding Light on Quantum Non-Classicality: \\ Witnessing Wigner Negativity with Free Electrons} 
}%

\author{Sunjun Hwang}%
\affiliation{Centre for Quantum Technologies,
National University of Singapore, Singapore 117543, Singapore}
\affiliation{School of Electrical and Electronic Engineering,
Nanyang Technological University, Singapore 639798, Singapore}
\author{Leong Chuan Kwek}%
\email{Contact author: kwekleongchuan@nus.edu.sg}
\affiliation{Centre for Quantum Technologies (CQT), National University of Singapore 117543, Singapore}
\affiliation{MajuLab, CNRS-UNS-NUS-NTU International Joint Research Unit, UMI 3654, Singapore}
\affiliation{School of Electrical and Electronic Engineering (EEE), Nanyang Technological University, 50 Nanyang Avenue, Singapore 639798}
\affiliation{National Institute of Education (NIE),
Nanyang Technological University, 1 Nanyang Walk, Singapore 637616}

\date{\today}

\begin{abstract}
Certifying quantum non-classicality like Wigner negativity typically demands full optical state tomography. This work introduces a noise-resilient framework for witnessing optical Wigner negativity using phase-resolved quantum photon-induced near-field electron microscopy (QPINEM). By mapping electron coherences directly to symmetrically ordered optical characteristic functions within a coupling-limited phase-space region, the protocol bypasses full optical reconstruction. A post-measurement witness matrix is constructed via Bochner's theorem from recorded electron spectra, where negative eigenvalues certify non-classicality. To ensure reliability under realistic experimental conditions, a noise-aware selection procedure decouples discovery from statistical validation. Benchmarked across various optical preparations—including vacuum-Fock superpositions, odd cat states, and GKP states—the protocol provides a statistically bounded pathway for probing quantum features with free electrons.
\end{abstract}

\maketitle


\noindent{\bf Introduction --} The ability to rigorously certify non-classical features of
optical fields is a cornerstone of modern quantum information
science and foundational quantum optics
\cite{Braunstein2005,Eisert2020}.
Among these signatures,
Wigner negativity---the failure of a quantum state's Wigner
quasiprobability distribution to remain non-negative everywhere
in phase space---serves as a definitive indicator of optical
non-classicality \cite{Kenfack2004} and an important resource
for continuous-variable quantum computation
\cite{Mari2012,Albarelli2018,TakagiZhuang2018,Walschaers2021}.
Traditionally, mapping and certifying Wigner negativity rely
on optical homodyne tomography followed by state reconstruction
\cite{Smithey1993,Lvovsky2001,LvovskyRaymer2009}, or phase-space
measurements using photon-number-resolving detectors
\cite{Banaszek1996,Banaszek1999,Laiho2010}.
However, as quantum states scale in complexity, full tomographic
reconstruction becomes increasingly demanding \cite{Mele2025},
motivating the development of alternative witnesses that can
certify non-classicality directly from selected measurements
without requiring complete state tomography
\cite{Mari2011,Kot2012,Chabaud2021,Zaw2024}.

Concurrently, the rapid maturation of free-electron quantum
optics \cite{GarciaDeAbajoDiGiulio2021,ruimy2025free}
has established ultrafast electron microscopy as a
powerful platform for probing optical fields with nanometer-scale
spatial resolution \cite{GarciaDeAbajo2010,Polman2019},
while coherent electron control has reached the attosecond
timescale \cite{Priebe2017,MorimotoBaum2018}.
Through photon-induced near-field electron microscopy (PINEM)
\cite{Barwick2009,Feist2015} and its quantum extension, QPINEM,
the coherent interaction between swift electrons and optical
fields enables the translation of optical quantum statistics
into energy shifts and coherences of the probing electron beam
\cite{Kfir2019,DiGiulio2019,BenHayun2021,Dahan2021}.
Recent theoretical advances in phase-resolved electron
spectroscopy have further shown how optical characteristic
functions can be recovered from reduced electron density
matrices within a coupling-limited phase-space region,
bridging free-electron dynamics and optical phase-space
structures \cite{Gorlach2024}.

Despite these developments, bridging free-electron readouts
with rigorous, noise-resilient criteria for optical Wigner
negativity poses a distinct challenge.
In realistic physical scenarios, counting statistics,
finite experimental budgets, and reconstruction biases can
obscure genuine non-classical signatures, leading to false
positives or missed detections when standard witnessing
strategies are applied naively \cite{Schwemmer2015}.
Reliable state characterization therefore requires explicit
uncertainty bounds \cite{ChristandlRenner2012,FaistRenner2016}. In particular, selecting and validating a witness using the
same noisy data can bias its inferred statistical significance
\cite{Cox1975}.

Here, we introduce a robust framework for free-electron
witnessing and noise-aware certification of optical Wigner
negativity using phase-resolved QPINEM readout.
By leveraging the mapping between electron coherences and
symmetrically ordered optical characteristic functions, our
protocol circumvents full optical-state tomography.
We establish a post-measurement witnessing scheme built upon
Bochner's theorem \cite{Bochner1933,Mari2011}, wherein sampling
geometries in phase space are systematically evaluated using
the same recorded electron spectra via a noise-aware selection
procedure that decouples discovery from statistical validation.
Using simulated benchmarks across diverse optical
preparations---ranging from vacuum--Fock preparations and
odd-cat states to finite-energy GKP states, photon-added
thermal states, and phase-diffused odd cats---we demonstrate
that our approach provides a reliable, statistically bounded
pathway to certify optical non-classicality using free-electron
probes, with finite-count guarantees conditional on the
calibrated state families and measurement model.


\bigskip

\noindent{\bf Result --} Consider a mono-energetic electron of energy $E_0$, initially uncorrelated with an unknown single-mode optical state $\hat\rho_p$, independently prepared for each probe electron. Within the paraxial, linear-dispersion approximation, quantum PINEM (QPINEM) coherently couples electron energy gain and loss to photon annihilation and creation, respectively, through a complex coupling amplitude $g_Q$ \cite{Kfir2019,BenHayun2021}. 
In the electron phase basis \cite{Gorlach2024}, tracing out the optical mode after QPINEM yields reduced electron density-matrix elements directly proportional to the optical characteristic function:
\begin{equation}
\langle\varphi|\hat\rho^{e}|\varphi'\rangle
=\frac{e^{i\operatorname{Im}(\xi_\varphi\xi_{\varphi'}^*)}}{2\pi}
\chi(\xi_\varphi-\xi_{\varphi'}).
\label{eq:electron_chi}
\end{equation}
Here, $|\varphi\rangle$ denotes an electron phase state, $\xi_\varphi=-g_Q^*e^{-i\varphi}$, and $\chi(\xi)=\operatorname{Tr}[\hat\rho_p\hat D(\xi)]$ is the symmetrically ordered optical characteristic function, with $\hat D(\xi)$ being the optical displacement operator \cite{Leonhardt1997,Mari2011}. Equation~(\ref{eq:electron_chi}) establishes that the electron coherences encode the optical characteristic function, up to a known phase factor. The detailed derivation is provided in End Matter~\ref{app:protocol}. Because the displacement arguments lie on a circle of radius $|g_Q|$, their differences span the disk $|\xi|\leq 2|g_Q|$ \cite{Gorlach2024}. Thus, the QPINEM coupling determines the accessible phase-space region, independently of the reference coupling strength $g_R$.

\begin{figure*}[t]
\centering
\includegraphics[width=0.85\textwidth]{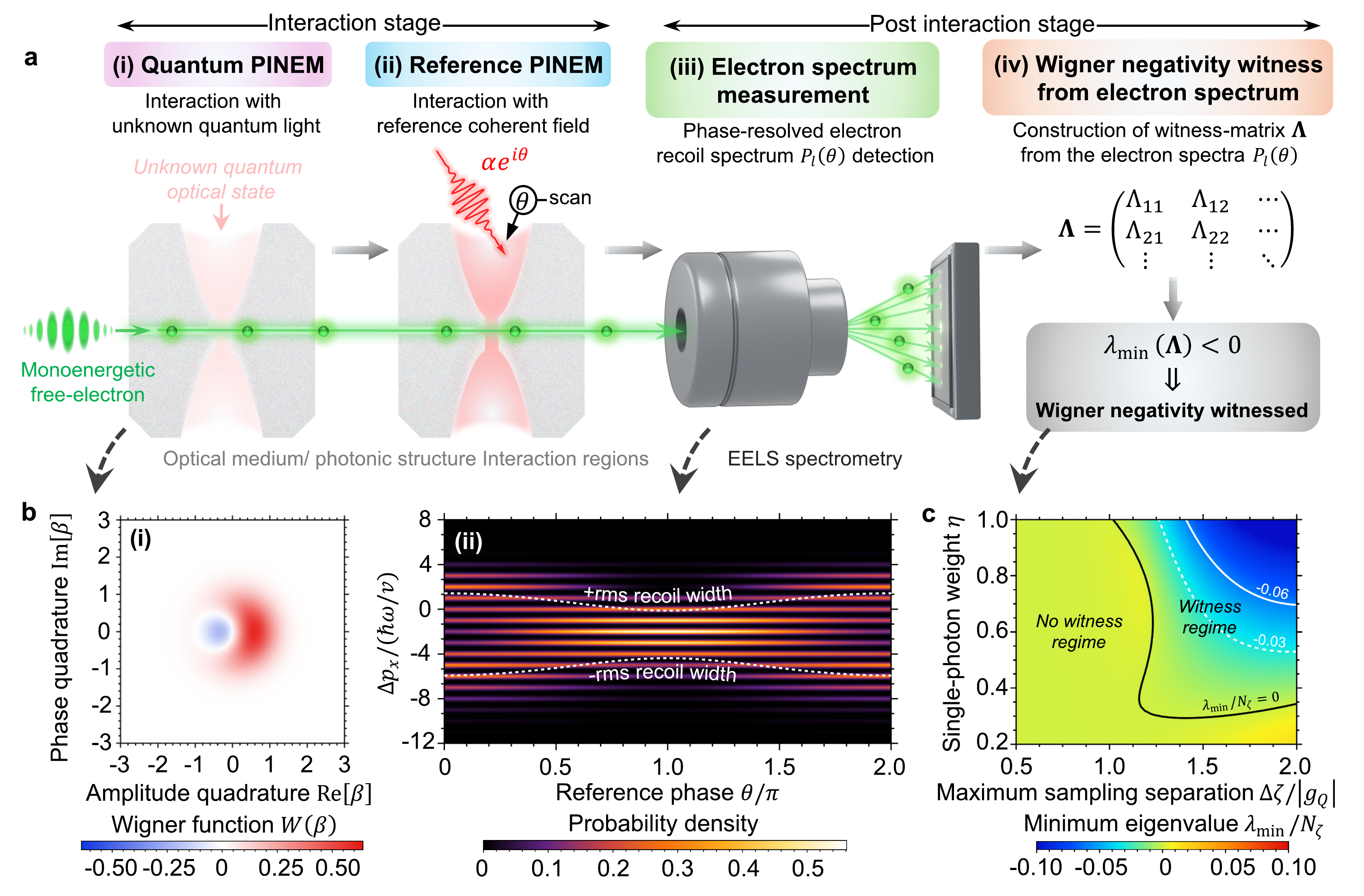}
\caption{\textbf{Free-electron protocol for witnessing optical Wigner negativity.} \textbf{(a)} A monoenergetic electron is sequentially (i) coupled to an unknown quantum optical state via quantum PINEM (coupling strength $g_Q$), (ii) modulated by a phase-controlled coherent reference field $\alpha e^{i\theta}$ via reference PINEM (coupling strength $g_R$), and (iii) analyzed by EELS to obtain the phase-resolved electron spectra $P_l(\theta)$. (iv) The measured spectra provide values of the optical characteristic function $\chi(\xi)$ at accessible phase-space points, from which the witness matrix $\Lambda_{jk}=\chi(\zeta_j-\zeta_k)$ is constructed. A negative eigenvalue of $\Lambda$ certifies Wigner negativity of the unknown optical state. \textbf{(b)} Vacuum--single-photon superposition $|\psi_\eta\rangle=\sqrt{1-\eta}|0\rangle+\sqrt{\eta}|1\rangle$, with single-photon weight $\eta=0.5$: (i) the corresponding Wigner function exhibiting negativity. (ii) The phase-resolved electron recoil distribution following the two PINEM interactions ($g_Q=g_R=1.5$). Sweeping $\theta$ converts electron coherences into phase-dependent populations, producing distinct spectra. White dashed curves indicate the mean recoil $\pm$ rms spread.
\textbf{(c)} Minimum eigenvalue of $\Lambda$, normalized by the number of sampling points $N_\zeta=9$, versus $\eta$ and the normalized maximum sampling separation $\Delta\zeta/|g_Q|$, where $\Delta\zeta=\max_{j,k}|\zeta_j-\zeta_k|$.
The sampling points form a center-plus-octagon geometry, with orientation optimized to minimize $\lambda_{\min}$. Negative (blue) regions certify Wigner negativity, whereas nonnegative regions are inconclusive.}
\label{fig:1}
\end{figure*}

\textit{From optical coherence to an electron witness.---}
Figure~\ref{fig:1}(a) illustrates a free-electron protocol for witnessing Wigner negativity in an unknown optical state with angular frequency $\omega$, combining free-electron quantum-optical readout \cite{Gorlach2024} with characteristic-function-based witnessing \cite{Mari2011,Zaw2024}. 

Following QPINEM, the electron interacts with a coherent reference field of controllable amplitude $\alpha$ and phase $\theta$ at the same frequency $\omega$. This reference-PINEM interaction, characterized by an effective
coupling strength $g_R$, coherently mixes the electron energy
sidebands \cite{Feist2015,Echternkamp2016,Priebe2017}. Assuming negligible dispersive propagation between the two interactions, the probability of detecting an electron at energy $E_0+l\hbar\omega$ is
\begin{equation}
P_l(\theta)=\sum_{m,q}A_{lm}^{(q)}
\rho^{e}_{m,m+q}e^{iq\theta},
\label{eq:electron_spectra}
\end{equation}
where $\rho^e_{mn}=\langle m|\hat\rho^e|n\rangle$ and $A_{lm}^{(q)}=J_{l-m}(2g_R)J_{l-m-q}(2g_R)$, with $J_n$ denoting the Bessel function of the first kind. 
Here, $q$ is the coherence order: $q=0$ describes populations, while $q\neq0$ labels coherences between sidebands separated by $q\hbar\omega$. Scanning $\theta$ converts these coherence orders into distinct Fourier harmonics of the measured populations. Fourier analysis and regularized inversion estimate electron coherences and hence accessible values of $\chi$ through Eq.~(\ref{eq:electron_chi}). The derivation is given in End Matter,~\ref{app:protocol}. 
Figure~\ref{fig:1}(b) illustrates this transfer: optical Wigner negativity can be inferred from the coherences encoded in the phase-resolved electron spectra.

To extract a signature of Wigner negativity, we select complex phase-space points $\{\zeta_j\}$ whose pairwise differences satisfy $|\zeta_j-\zeta_k|\leq 2|g_Q|$ for all $j,k$. By Bochner's theorem, a nonnegative Wigner function $W(\beta)\geq0$ for all $\beta$ requires the matrix with elements $\chi(\zeta_j-\zeta_k)$ to be positive-semidefinite for any choice of sampling points \cite{Bochner1933,Mari2011}. Therefore, the condition
\begin{equation}
\Lambda_{jk}=\chi(\zeta_j-\zeta_k),\qquad
\lambda_{\min}(\Lambda)<0
\label{eq:negativity_witness}
\end{equation}
certifies Wigner negativity without reconstructing $W$ throughout phase space. Here, $\lambda_{\min}(\Lambda)$ denotes the smallest eigenvalue of the witness matrix $\Lambda$. The converse, however, does not hold: a nonnegative minimum eigenvalue, $\lambda_{\min}(\Lambda)\geq0$, for a given sampling geometry does not certify Wigner nonnegativity. Figure~\ref{fig:1}(c) illustrates this distinction: the same Wigner-negative state may be detected or missed depending on the sampling separation.

\begin{figure*}[t]
\centering
\includegraphics[width=0.90\textwidth]{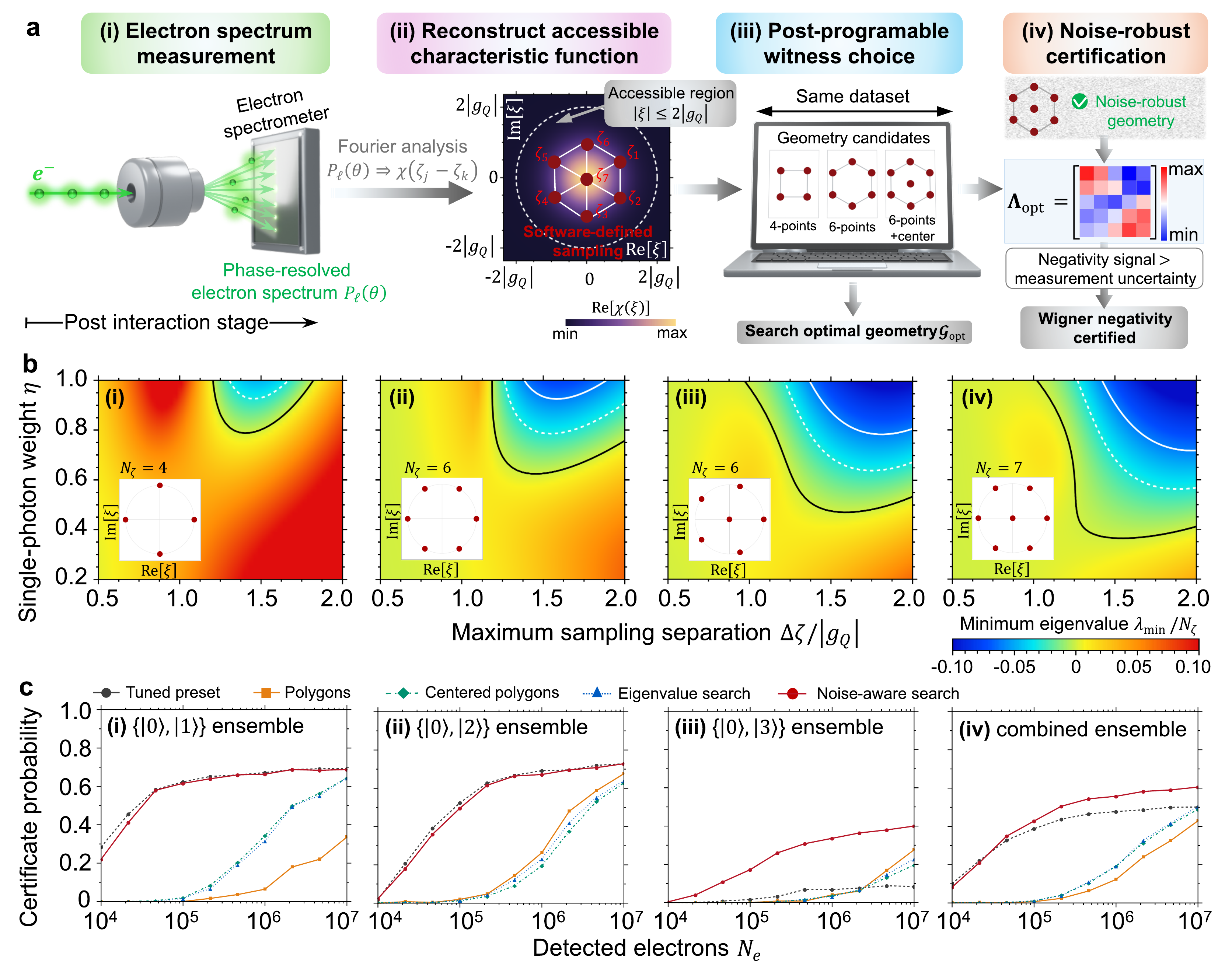}
\caption{\textbf{Post-measurement witness optimization.} \textbf{(a)} (i,ii) Fourier analysis and inversion of the electron spectra estimate $\chi(\xi)$ within $|\xi|\leq 2|g_Q|$. Sampling points $\{\zeta_j\}$ are selected in software such that $|\zeta_j-\zeta_k|\leq 2|g_Q|$ for all pairs. (iii) Candidate geometries and their minimum-eigenvalue vectors are evaluated using discovery spectra, without further data acquisition. (iv) Certification requires a negative upper confidence bound after accounting for reconstruction and counting uncertainties.
\textbf{(b)} Normalized minimum eigenvalue $\lambda_{\min}/N_\zeta$ for the vacuum--single-photon superposition under different sampling-point geometries (insets), each with optimized orientation.
\textbf{(c)} Certification probability versus detected electrons $N_e$ for partially coherent vacuum--$|s\rangle$ states ($s=1,2,3$) and the equally weighted average of their ensembles. The $|s\rangle$ weight $\eta\in(0.5,1)$, residual coherence $\gamma\in[0,1]$, and relative phase $\phi$ are randomized. Five geometry strategies (legend) are compared using training-tuned reconstruction and calibrated witness-error bounds. Certification probability is the fraction of 250 simulated trials per state family that certify negativity at 95 $\%$ confidence. In each trial, 30$\%$ of electrons are used for witness selection and the remaining 70$\%$ for independent validation ($g_Q=g_R=1.5$).}
\label{fig:2}
\end{figure*}

\textit{Post-measurement witnessing from electron spectra.---}
Figure~\ref{fig:2}(a) illustrates how the phase-resolved electron spectra can be used to construct different Wigner-negativity witnesses after measurement. A sampling geometry $\mathcal G=\{\zeta_j\}$ determines which characteristic-function values $\chi(\zeta_j-\zeta_k)$, reconstructed from the electron data, enter the witness matrix. Changing the number, arrangement, or orientation of the sampling points can alter the minimum eigenvalue and, consequently, the sensitivity to Wigner negativity, without requiring additional electron measurements. Figure~\ref{fig:2}(b) illustrates how the number and arrangement
of sampling points affect the ideal detection region; a larger matrix does not by itself guarantee greater
sensitivity.

To quantitatively compare different sampling geometries, we introduce the normalized witness matrix $B_{\mathcal G}=\Lambda_{\mathcal G}/N_\zeta$, where $N_\zeta$ is the number of sampling points. For each geometry, the eigenvector $\bm v$ associated with $\lambda_{\min}(B_{\mathcal G})$ determines the combination of characteristic-function values yielding the smallest scalar witness $t_{\mathcal G}=\bm v^\dagger B_{\mathcal G}\bm v$. Increasing $N_\zeta$ provides additional coefficients for combining the sampled characteristic-function values, potentially revealing negativity missed by a smaller geometry.

\begin{figure*}[t]
\centering
\includegraphics[width=0.93\textwidth]{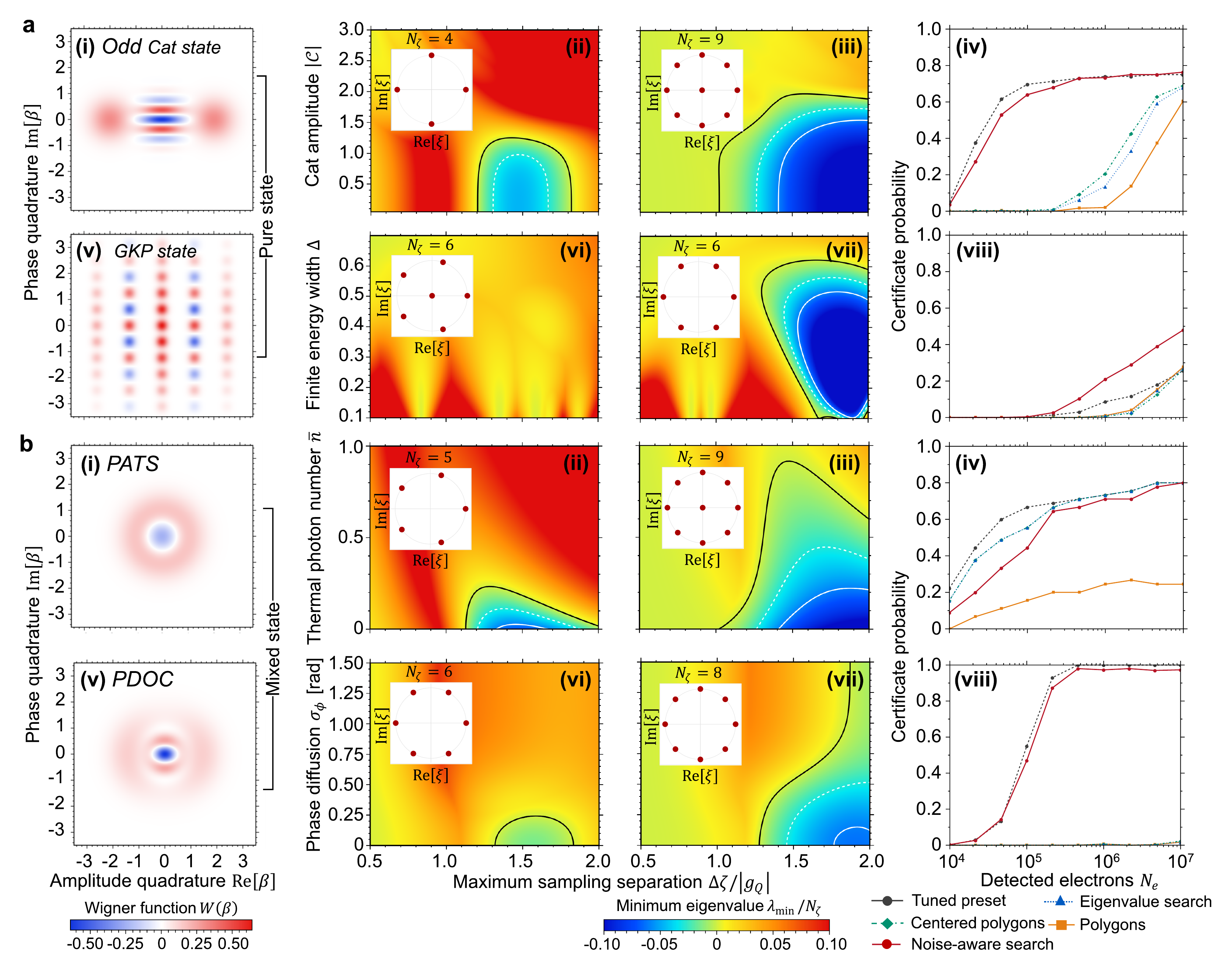}
\caption{\textbf{Wigner-negativity witnessing across diverse optical states.}
\textbf{(a)} Pure states: (i) Wigner function of an odd Schr{\"o}dinger cat state $|\psi_-(\mathcal C)\rangle\propto|\mathcal C\rangle-|-\mathcal C\rangle$, where $|\pm\mathcal C\rangle$ are coherent states with dimensionless complex amplitudes $\pm\mathcal C$; (ii,iii) $\lambda_{\min}/N_\zeta$ versus $|\mathcal C|$ and $\Delta\zeta/|g_Q|$ for different sampling geometries (insets); (iv) certification probability versus detected electrons $N_e$. (v--viii) Corresponding results for a finite-energy GKP state, varying its Gaussian peak-width parameter $\Delta$. At fixed $N_\zeta=6$, the regular hexagon in (vii) reveals negativity missed by the center-plus-pentagon in (vi), showing that a central point does not by itself ensure greater witnessing sensitivity.
\textbf{(b)} Mixed states: (i--iv) single-photon-added thermal states (PATS), varying the mean photon number $\bar n$ of the thermal seed before photon addition; (v--viii) phase-diffused odd-cat states (PDOC), varying the phase-diffusion width $\sigma_\phi$ at fixed $|\mathcal C|=1.5$. The eigenvalue maps use $g_Q=1.5$. Finite-count benchmarks follow the training-tuned procedure of Fig.~2(c), with family-calibrated witness-error bounds and $g_Q=1.5$, $g_R=4$. Differences reflect how each state's phase-space structure is captured by accessible characteristic-function values, while finite-count certification additionally requires the negative witness signal to exceed reconstruction and counting uncertainties. The benchmark states and their parameters are defined in the End Matter. Detailed state definitions and calibration parameters are provided in the Supplemental Material.
}
\label{fig:3}
\end{figure*}

With finite electron counts, however, greater sampling freedom does not necessarily improve witnessing reliability. Counting fluctuations in the measured electron energy spectra, together with errors from reconstructing electron coherences, can produce spuriously negative witness values. Searching over many candidate geometries may further amplify this effect. We therefore separate witness selection from its statistical validation using independent electron measurements \cite{Cox1975}. After fixing the reconstruction settings through independent training, a subset of measurements (discovery data) determines the minimum-eigenvalue vector $\bm v_{\mathcal G}$ for each candidate geometry. Rather than selecting the most negative reconstructed eigenvalue alone, we optimize
\begin{equation}
\mathcal G_* = \operatorname*{arg\,min}_{\mathcal G}
\left[t_{\mathcal G,\mathrm{rec}}^{(d)}
+b_{\mathcal G}+\epsilon_{\mathcal G}^{(d)}\right].
\label{eq:geometry_optimization}
\end{equation}
Here, $t_{\mathcal G,\mathrm{rec}}^{(d)}=\bm v_{\mathcal G}^\dagger B_{\mathcal G,\mathrm{rec}}^{(d)}\bm v_{\mathcal G}$ is the estimated witness value, while $b_{\mathcal G}$ and $\epsilon_{\mathcal G}^{(d)}$ account for reconstruction bias and counting uncertainty, respectively. This noise-aware selection favors geometries with the lowest uncertainty-adjusted witness values.

To certify Wigner negativity, the selected geometry and eigenvector are held fixed and evaluated using independent electron measurements. Each recorded reference-phase setting and electron energy outcome contributes to the selected witness through a predetermined measurement weight, analogous to sampling-function approaches in quantum optics \cite{Fiurasek2013}. The resulting counting uncertainty is bounded using the empirical Bernstein inequality \cite{MaurerPontil2009}. Certification requires
\begin{equation}
t_{\mathcal G_*,\mathrm{rec}}^{(v)}
+b_{\mathcal G_*}
+\epsilon_{\mathcal G_*}^{(v)}(\delta)
<0.
\label{eq:statistical_certification}
\end{equation}
Here, $\epsilon_{\mathcal G_*}^{(v)}(\delta)$ denotes a statistical error bound with failure probability $\delta$. Assuming independent validation measurements, the probability of falsely certifying Wigner negativity is at most $\delta$ for Wigner-nonnegative states covered by the reconstruction-error calibration, at a fixed electron budget (see End Matter,~\ref{app:optimization}).

To assess the witnessing performance for unknown optical preparations, Fig.~\ref{fig:2}(c) applies the protocol to simulated electron measurements of states randomly sampled within the Fock subspaces $\operatorname{span}\{|0\rangle,|s\rangle\}$ ($s=1,2,3$). The tuned preset performs comparably to noise-aware selection for $s=1,2$, whereas the latter offers a clear advantage for $s=3$, where additional radial oscillations and higher-order angular dependence of the characteristic function increase sensitivity to the sampling geometry. This advantage also carries over to the combined ensemble. Increasing the electron count reduces statistical uncertainty but does not necessarily eliminate reconstruction bias or compensate for a poorly matched sampling geometry.

Figure~\ref{fig:3} extends our analysis to pure odd-cat and finite-energy GKP states \cite{Gottesman2001, Walschaers2021}, and mixed photon-added thermal \cite{Zavatta2007} and phase-diffused odd-cat states. The ideal eigenvalue maps reveal the dependence of witnessing sensitivity on the sampling geometry, while finite-count certification is particularly challenging for GKP states, highlighting the importance of accessible characteristic-function information and reconstruction accuracy.

The mixed-state benchmarks further demonstrate the versatility of electron-based witnessing. Photon-added thermal states can be certified from phase-independent electron spectra, whereas phase-diffused odd-cat states remain detectable despite suppressed phase-sensitive coherences. Across these states, noise-aware selection generally improves certification over direct eigenvalue optimization by accounting for measurement uncertainties. These benchmarks assume calibrated state families and measurement conditions, with GKP calibration restricted to a finite parameter catalogue (see End Matter and Supplemental Material).


\noindent{\bf Conclusion --} Our study establishes a free-electron approach to witnessing optical Wigner negativity by combining quantum-optical electron readout \cite{Gorlach2024} with characteristic-function witnesses \cite{Mari2011,Zaw2024}. Through quantum PINEM and phase-resolved EELS, information about an unknown optical state is transferred to electron coherences and retrieved without direct photon detection or full optical-state tomography. Crucially, the same acquired electron spectra enable post-measurement optimization of witness sampling geometries and independent, uncertainty-bounded certification without additional measurements. Our numerical benchmarks across pure and mixed optical states demonstrate the importance of accounting for reconstruction and counting uncertainties.

Our study also highlights a practical challenge: witnessing certain optical states may require strong quantum-PINEM coupling to access the relevant characteristic-function region. Achieving this regime remains an engineering challenge, with attainable coupling strengths subject to material- and geometry-dependent physical bounds \cite{Xie2025}. Moreover, the present finite-count guarantees assume independent optical preparations and calibrated reconstruction errors. A possible route toward extending these guarantees to broader classes of unknown optical states is to combine experimentally characterized electron measurement operators with reconstruction-error bounds uniform over the enlarged class, potentially under suitable optical-energy constraints
\cite{Lundeen2009,Gandhari2024,Mele2025}. Experimental implementations must additionally characterize finite energy and phase resolution, coupling fluctuations, and electron propagation effects.

Strong-coupling nanophotonic designs and guided-electron schemes provide promising directions for experimental implementation \cite{Dmello2023,Karnieli2024Strong,Xie2025}. Relevant experimental advances include coherent electron
modulation by driven optical cavities \cite{Kfir2020,Wang2020},
extended phase-matched electron--light interactions \cite{Dahan2020}, and cavity-mediated electron--photon pair generation \cite{Feist2022}. Further developments include reconstruction-error bounds
applicable to broader optical-state classes and electron readout of multimode correlations. Together, these developments could extend free-electron spectroscopy beyond characterizing optical excitations
toward statistically certified witnessing of Wigner negativity in confined nanophotonic modes, leveraging the spatial and temporal selectivity of free-electron probes.

\begin{acknowledgments}

\end{acknowledgments}

\bibliography{apssamp}

\clearpage
\section*{End Matter}
\appendix

\section{Free-electron witnessing protocol}
\label{app:protocol}

\textit{Optical information transfer.---} 
We consider a monoenergetic free electron and an unknown single-mode quantum optical state, initially uncorrelated and described by the density operator $\hat{\rho}_{\mathrm{in}}=|0\rangle\langle0|\otimes\hat{\rho}_p$. Within the paraxial, linear-dispersion approximation \cite{BenHayun2021}, the electron is described by an equally spaced energy ladder $E_m=E_0+m\hbar\omega$, where $\omega$ is the angular frequency of the interacting optical mode. The corresponding energy-lowering operator satisfies $\hat b|m\rangle=|m-1\rangle$ and $[\hat b,\hat b^\dagger]=0$. Unlike a bosonic annihilation operator, $\hat b$ translates the energy index without a photon-number-dependent factor. The quantum-PINEM interaction is described by the scattering operator $\hat S_Q=\exp(g_Q\hat b^\dagger\hat a-g_Q^*\hat b\hat a^\dagger)$ \cite{Kfir2019,BenHayun2021,Gorlach2024}.

We introduce the electron phase basis $|\varphi\rangle=(2\pi)^{-1/2}\sum_m e^{-im\varphi}|m\rangle$, whose states satisfy $\hat b|\varphi\rangle=e^{-i\varphi}|\varphi\rangle$. In this basis, the quantum-PINEM (QPINEM) interaction becomes a phase-dependent displacement of the optical mode,
\begin{equation}
\hat S_Q=\int_0^{2\pi}d\varphi\,
|\varphi\rangle\langle\varphi|\otimes\hat D(\xi_\varphi).
\label{eq:em_conditional_displacement}
\end{equation}
Thus, different electron phase components induce different optical displacements. Their relative coherence after the interaction contains information about the unknown optical state. Here $\hat D(\xi)=\exp(\xi\hat a^\dagger-\xi^*\hat a)$ and
$\xi_\varphi=-g_Q^*e^{-i\varphi}$. Tracing out the photonic degree of freedom from the post-interaction state $\hat{\rho}_f=\hat S_Q\hat{\rho}_{\mathrm{in}}\hat S_Q^\dagger$ yields the reduced electron state $\hat{\rho}^e=\operatorname{Tr}_p(\hat{\rho}_f)$. Its phase-basis matrix elements are $\langle\varphi|\hat{\rho}^e|\varphi'\rangle=(2\pi)^{-1}\operatorname{Tr}_p[\hat{\rho}_p\hat D^\dagger(\xi_{\varphi'})\hat D(\xi_\varphi)]$. Using the identity $\hat D^\dagger(\xi')\hat D(\xi)=e^{i\operatorname{Im}(\xi\xi'^*)}\hat D(\xi-\xi')$ then yields Eq.~(\ref{eq:electron_chi}). The displacement arguments lie on a circle of radius $|g_Q|$;
their differences therefore fill $|\xi|\leq2|g_Q|$.
The boundary is set by the quantum coupling $g_Q$: increasing the reference strength can improve readout but does not enlarge this disk. Accessibility alone does not guarantee equal reconstruction accuracy at every point.

\textit{Reference-phase readout.---}
Assuming negligible electron dispersion between the two PINEM interactions, the phase-controlled reference PINEM is described by $\hat S_R(\theta)=\exp[g_R(e^{i\theta}\hat b^\dagger-e^{-i\theta}\hat b)]$, whose matrix elements in the electron energy basis are $\langle l|\hat S_R(\theta)|m\rangle=J_{l-m}(2g_R)e^{i(l-m)\theta}$. The resulting population of the $l$-th electron sideband is $P_l(\theta)=\sum_{m,n}\rho^e_{mn}\langle l|\hat S_R(\theta)|m\rangle\langle n|\hat S_R^\dagger(\theta)|l\rangle$. Substituting the matrix elements and setting $q=n-m$ yields Eq.~(\ref{eq:electron_spectra}), where $A_{lm}^{(q)}=J_{l-m}(2g_R)J_{l-m-q}(2g_R)$. The reference mixes energy sidebands coherently, converting off-diagonal electron coherences into phase-dependent populations. Fourier analysis of $P_l(\theta)$ then isolates the coherence order $q$, giving $\widetilde P_l^{(q)}=(2\pi)^{-1}\int_0^{2\pi}P_l(\theta)e^{-iq\theta}d\theta$. The Fourier coefficients are related to the electron density-matrix elements through
\begin{equation}
\widetilde{\bm P}^{(q)}=A^{(q)}\bm\rho^{(q)},\qquad
\bm\rho_{\mathrm{rec}}^{(q)}=(A^{(q)})_\tau^+\widetilde{\bm P}_{\mathrm{rec}}^{(q)}.
\label{eq:em_inverse}
\end{equation}
Here, $(\bm\rho^{(q)})_m=\rho^e_{m,m+q}$, $A^{(q)}$ is known from the reference interaction, and $\bm\rho_{\mathrm{rec}}^{(q)}$ denotes the reconstructed coherences. The truncated pseudoinverse $(A^{(q)})_\tau^+$ retains only singular values above a relative threshold $\tau$, reducing noise amplification at the cost of reconstruction bias \cite{GolubVanLoan2013}. Transforming the reconstructed electron coherences into the phase basis and using Eq.~(\ref{eq:electron_chi}) yields
\begin{equation}
\chi_{\mathrm{rec}}(\xi)=e^{-i\Phi}
\sum_{m,n}\rho_{\mathrm{rec},mn}^{e}
e^{im\varphi-in\varphi'}.
\label{eq:em_chi_reconstruction}
\end{equation}
Here, $\xi=\xi_\varphi-\xi_{\varphi'}$, $\Phi=\operatorname{Im}(\xi_\varphi\xi_{\varphi'}^*)$, and $\rho_{\mathrm{rec},mn}^{e}$ denotes the reconstructed electron density-matrix elements. The sum runs over the retained electron sidebands and coherence orders. This reconstructs selected values of the optical characteristic function without recovering the entire Wigner distribution. The reconstructed characteristic function is therefore an estimate: finite phase sampling, limited sideband detection, and regularized inversion can cause systematic deviations from the exact $\chi$ \cite{Gorlach2024, GolubVanLoan2013}. Finite phase sampling can likewise limit reconstruction accuracy in optical homodyne tomography \cite{LeonhardtMunroe1996}. We account for the errors in the present electron readout through
the calibrated reconstruction bounds below.

\textit{Negativity criterion.---}
Bochner's theorem \cite{Bochner1933,Mari2011,Zaw2024} implies that the symmetrically ordered characteristic function of a nonnegative Wigner function generates positive-semidefinite matrices for arbitrary sampling points. Substituting $\chi(\xi)=\int d^2\beta\,W(\beta)e^{\beta^*\xi-\beta\xi^*}$ into $\Lambda_{jk}=\chi(\zeta_j-\zeta_k)$ yields, for any complex vector $\bm c$,
\begin{equation}
\bm c^\dagger\Lambda\bm c
=\int d^2\beta\,W(\beta)
\left|\sum_j c_j e^{\beta\zeta_j^*-\beta^*\zeta_j}\right|^2.
\label{eq:em_bochner}
\end{equation}
Since the integral is nonnegative for $W(\beta)\geq0$, a negative eigenvalue of $\Lambda$ certifies Wigner negativity, establishing Eq.~(\ref{eq:negativity_witness}). Only finitely many accessible values of $\chi$ are needed to test a chosen matrix; a full phase-space reconstruction is therefore unnecessary.

\section{Post-measurement optimization and certification}
\label{app:optimization}

\textit{Training and witness selection.---}
For a geometry $\mathcal G$ containing $N_\zeta$ points, we use the normalized witness matrix $B_{\mathcal G}=\Lambda_{\mathcal G}/N_\zeta$. This allows geometries of different sizes to be compared without changing the sign of any eigenvalue. Since the witness matrix depends on $\chi(\zeta_j-\zeta_k)$, the separation between any two sampling points must satisfy $|\zeta_j-\zeta_k|\leq 2|g_Q|$. Within this accessible range, different sampling geometries can be explored numerically using the same recorded electron spectra, without additional measurements. We consider regular polygons with or without a central point, allowing their sizes and orientations to vary. The reconstruction settings and fixed-geometry benchmark are selected beforehand using independent training simulations, without consulting the test measurements. To avoid selecting and testing a witness using the same statistical fluctuations, the detected electrons are divided into discovery (30\%) and independent validation (70\%) datasets \cite{Cox1975}. For each candidate geometry $\mathcal G$, the discovery data determine the normalized eigenvector $\bm v_{\mathcal G}$ associated with the smallest eigenvalue of the reconstructed witness matrix. This vector identifies the direction of the strongest apparent negativity signal for that geometry, although the signal may include statistical noise. Since the reconstruction is linear in the measured electron populations, the corresponding witness value can be evaluated as an average over individual electron measurements, analogous to sampling-function approaches in quantum-optical measurements \cite{Mari2011,Fiurasek2013}:
\begin{equation}
t_{\mathcal G,\mathrm{rec}}^{(s)}
=\bm v_{\mathcal G}^\dagger B_{\mathcal G,\mathrm{rec}}^{(s)}\bm v_{\mathcal G}
=\frac{1}{N_\zeta}+\frac{1}{n_s}\sum_{i=1}^{n_s}w_{\mathcal G}(X_i^{(s)}).
\label{eq:em_event_score}
\end{equation}
Here, $s=d,v$ denotes discovery or validation, $n_s$ is the corresponding electron count, and $B_{\mathcal G,\mathrm{rec}}^{(s)}$ is the normalized matrix reconstructed from that dataset. Each outcome $X_i^{(s)}$ specifies a reference phase and a detected electron sideband; its real-valued score $w_{\mathcal G}(X_i^{(s)})$ is fixed once the geometry and eigenvector are chosen. The same scoring rule is then applied to independent validation electrons, with statistical uncertainty bounded using the score variance and its known range \cite{MaurerPontil2009}.

\textit{Reconstruction and counting uncertainties.---}
Reliable certification must distinguish two sources of uncertainty. Reconstruction bias comes from approximations such as finite electron-energy windows and regularized inversion; unlike counting fluctuations, it does not necessarily decrease when more electrons are detected. For the chosen geometry $\mathcal G$ and eigenvector $\bm v_{\mathcal G}$, a nonnegative allowance $b_{\mathcal G}$ bounds how much the ideal witness may exceed the mean reconstructed value. It is calibrated over a specified optical-state family, including the effects of regularization, phase sampling, and truncation, without using the unknown test state's parameters. The mathematical construction of this one-sided allowance is given in the Supplemental Material.

The counting uncertainty is determined from the individual electron measurement scores. For $n$ independent measurements with sample variance $S^2_{\mathcal G}$ and known score range $R_{\mathcal G}$, the empirical Bernstein inequality \cite{MaurerPontil2009} gives
\begin{equation}
\epsilon_{\mathcal G}(n,\delta)
=\sqrt{\frac{2S^2_{\mathcal G}\ln(2/\delta)}{n}}
+\frac{7R_{\mathcal G}\ln(2/\delta)}{3(n-1)}.
\label{eq:em_bernstein}
\end{equation}
Here, $\delta$ is the allowed failure probability, and $\epsilon_{\mathcal G}$ bounds the statistical deviation of the estimated witness value for a fixed witness. During discovery, the geometry optimization in Eq.~(\ref{eq:geometry_optimization}) favors a negative witness signal that is strong relative to both reconstruction and counting uncertainties, rather than merely the most negative reconstructed eigenvalue. This search does not itself provide a confidence guarantee because the eigenvector was fitted to the same discovery data. Independent validation of the fixed geometry and eigenvector enables the statistically justified certification criterion in Eq.~(\ref{eq:statistical_certification}).

\textit{Independent certification and scope.---}
Once the sampling geometry $\mathcal G_*$ and witness eigenvector $\bm v_{\mathcal G_*}$ are selected, both remain fixed during validation. The independent electron measurements then provide an estimate of the same witness, allowing its statistical uncertainty to be bounded without selection bias. Combining this uncertainty with the calibrated reconstruction-error allowance gives the upper confidence bound $U_\delta$, defined by the left-hand side of Eq.~(\ref{eq:statistical_certification}). If $U_\delta<0$, a nonnegative-Wigner model is rejected with false-certification probability at most $\delta$, under the stated assumptions. We use $\delta=0.05$, corresponding to 95\% confidence. A failure to obtain a negative upper bound is inconclusive, not evidence of a positive Wigner function.

The finite-count guarantees assume calibrated optical-state families: vacuum--Fock states supported on $\operatorname{span}\{|0\rangle,|s\rangle\}$ ($s=1,2,3$); odd cats $|\psi_-\rangle\propto|\mathcal C\rangle-|-\mathcal C\rangle$ and their phase-diffused counterparts; photon-added thermal states $\hat\rho_{\mathrm{PATS}}\propto\hat a^\dagger\hat\rho_{\mathrm{th}}(\bar n)\hat a$; and Gaussian-enveloped GKP states characterized by peak width $\Delta$. Calibration covers the specified continuous parameter ranges, except for GKP states, which are restricted to a finite width--orientation catalogue. Complete state definitions and calibration details are provided in the Supplemental Material. Thus, the ideal negativity criterion is state-independent, whereas the finite-count guarantees remain conditional on the calibration assumptions.

\end{document}